\documentclass[9pt,shortpaper,twoside,web]{IEEEtran}
\usepackage[noadjust]{cite}
\usepackage{amsmath,amssymb,amsfonts}
\usepackage{algorithmic}
\usepackage{graphicx}
\usepackage{textcomp}
\usepackage{verbatim}
\usepackage[table]{xcolor}
\usepackage{soul}
\usepackage{makecell}
\usepackage{amsmath}
\usepackage{subcaption}
\usepackage{dblfloatfix}

\def\BibTeX{{\rm B\kern-.05em{\sc i\kern-.025em b}\kern-.08em
		T\kern-.1667em\lower.7ex\hbox{E}\kern-.125emX}}

\begin{document}	
	\title{Hybrid Pass Transistor Logic with Ambipolar Transistors}
	\author{Xuan Hu, \IEEEmembership{Student Member, IEEE}, Amy S. Abraham, Jean Anne C. Incorvia, \IEEEmembership{Member, IEEE}, and \\Joseph S. Friedman, \IEEEmembership{Senior Member, IEEE}
		\thanks{Manuscript submitted July XXX, 2020.}
		\thanks{X. Hu, A. S. Abraham, and J. S. Friedman are with the Department of Electrical and Computer Engineering, The University of Texas at Dallas, Richardson, TX 75080 USA (email: joseph.friedman@utdallas.edu).}
		\thanks{J. A. C. Incorvia is with the Department of Electrical and Computer Engineering, The University of Texas at Austin, Austin, TX 78712 USA.}

}
	\maketitle

	\vspace*{-2em}\begin{abstract}	
		In comparison to the conventional complementary pull-up and pull-down logic structure, the pass transistor logic (PTL) family reduces the number of transistors required to perform logic functions, thereby reducing both area and power consumption. However, this logic family requires inter-stage inverters to ensure signal integrity in cascaded logic circuits, and inverters must be used to provide each logical input signal in its complementary form. These inverters and complementary signals increase the device count and significantly degrade overall system efficiency.\par
		Dual-gate ambipolar field-effect transistors natively provide a single-transistor XNOR operation and permit highly-efficient and compact circuits due to their ambipolar capabilities. Similar to PTL, logic circuits based on ambipolar field-effect transistors require complementary signals. Therefore, numerous inverters are required, with significant energy and area costs.\par	
		Ambipolar field-effect transistors are a natural match for PTL, as hybrid ambipolar-PTL circuits can simultaneously use these inverters to satisfy their necessity in both PTL and ambipolar circuits. We therefore propose a new hybrid ambipolar-PTL logic family that exploits the compact logic of PTL and the ambipolar capabilities of ambipolar field-effect transistors. Novel hybrid ambipolar-PTL circuits were designed and simulated in SPICE, demonstrating strong signal integrity along with the efficiency advantages of using the required inverters to simultaneously satisfy the requirements of PTL and ambipolar circuits. In comparison to the ambipolar field-effect transistors in the conventional CMOS logic structure, our hybrid full adder circuit can reduce propagation delay by 47\%, energy consumption by 88\%, energy-delay product by a factor of 9, and area-energy-delay product by a factor of 20. \par
	\end{abstract}
	
	\begin{IEEEkeywords}
		Ambipolarity, Transmission Gate Logic, Pass Transistor Logic (PTL), Carbon Nanotube, Ambipolar Logic
	\end{IEEEkeywords}
	
	\section{Introduction}
	
	As technology scaling becomes increasingly challenging and expensive, novel device switching phenomena have the potential to revolutionize computing beyond conventional unipolar field-effect transistors (FETs) with a fixed charge polarity. In particular, dual-gate ambipolar FETs (DG-A-FETs) enable efficient computing through dynamic switching between \textit{n}- and \textit{p}-type \cite{Lin2004b, Lin2005, Hu2017a, Hu2017n,  Pan2018, Bucella2016, Zhang2013c, Zhang2014, Hu2019BookChapter}. Such DG-A-FETs can be implemented with a variety of materials that exhibit ambipolar transport, including carbon nanotubes (CNTs), silicon nanowires, graphene nanoribbons, and transition metal dichalcogenides \cite{Yin2017b, Zhang2012, Kim2017b, Ortiz2018, Tian2010,Yu2018, Baeg2017}. Though this paper considers circuits designed with DG-A-FETs with CNTs (DG-A-CNTFETs) due to the existence of an effective SPICE-compatible device model, the general conclusions are expected to also apply to DG-A-FETs based on other ambipolar materials.\par
	As illustrated in Fig. \ref{device}, the polarity gate (PG) voltage in a DG-A-FET determines the dominant carrier polarity while the control gate (CG) voltage modulates the current. Ambipolar transistors have been used in circuits based on the conventional complementary pull-up and pull-down logic family, enabling increased logical expressivity in compact circuits \cite{Ben-Jamaa2011b, Resta2016}. In particular, the conventional complementary pull-up and pull-down logic family enables XOR and XNOR gates with only four transistors. However, these gates require complementary input signals, necessitating the use of a two-transistor inverter for each input. These additional inverters significantly increase the area and energy consumption, thereby reducing the benefits of using ambipolar transistors.\par

	\begin{figure}[t]
		\vspace*{-0.5em}
		\centering
		\includegraphics[trim=25 215 25 215 ,clip, width=0.48\textwidth]{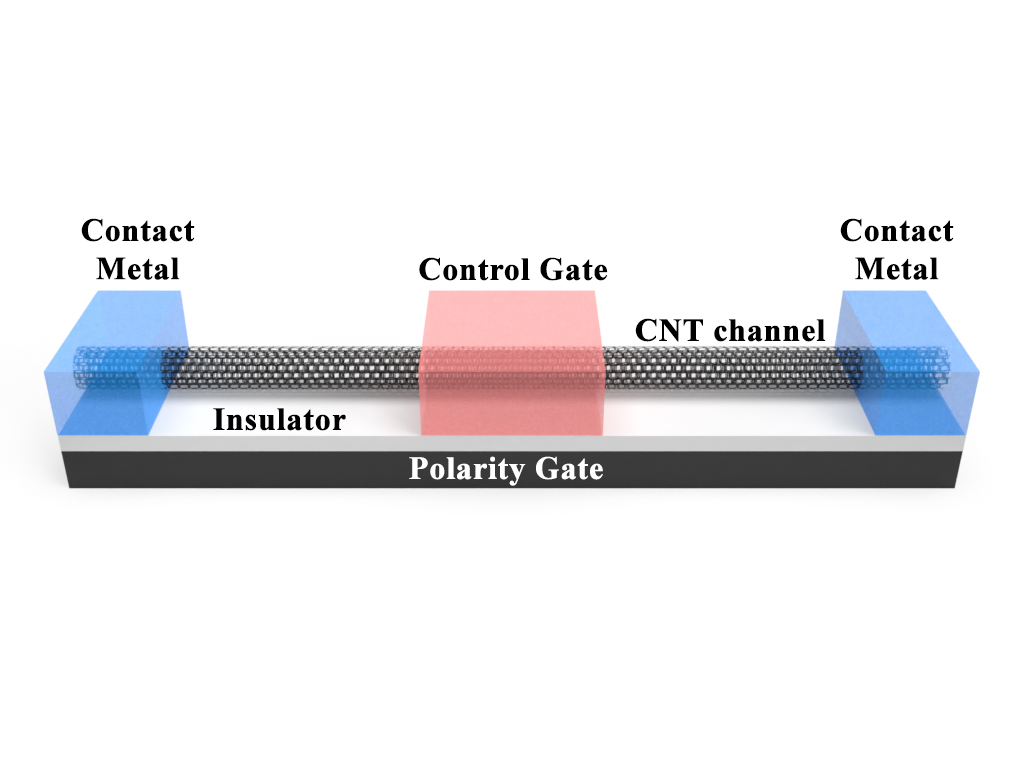}
		\vspace{-0.8em}
		\caption{Cross-section of a DG-A-FET with an ambipolar CNT channel.}
		\vspace{-1.5em}
		\label{device}
	\end{figure}

	\begin{figure*}[!b]
		\centering
		\includegraphics[width=0.85\textwidth]{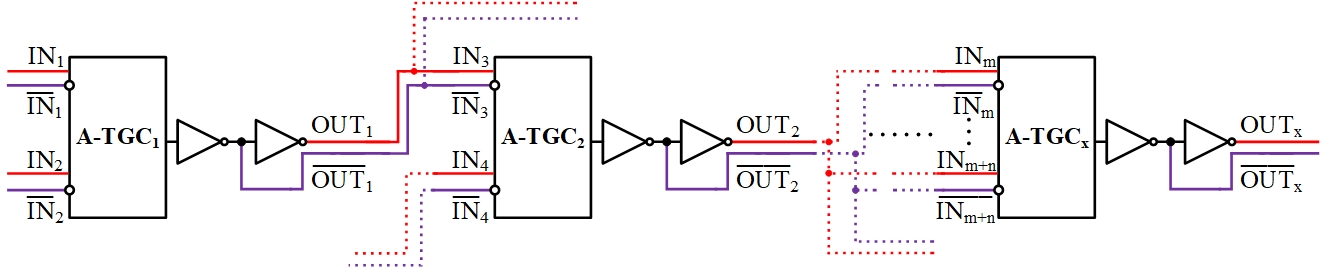}
		\setcounter{figure}{2}
		\caption{Block diagram of the Type I hybrid A-PTL architecture illustrating the alternation between A-TGCs and buffering inverters.}
		\vspace{-1.5em}
		\label{PTL_Diagram}
	\end{figure*}
	
	Pass transistor logic (PTL) is a well-established logic family that is designed to reduce the energy and area of logical computing systems by reducing the number of transistors required to perform logical functions \cite{Yano1996, Radhakrishnan1985, Yano1990}. However, its broad application in large-scale circuits is limited by the need for inverters to provide the external power required for signal integrity \cite{VonNeumann1993b}.\par

	As both DG-A-FET logic and the PTL family require additional inverters for proper operation, this paper proposes a hybrid A-PTL family that simultaneously uses the inverters to satisfy the requirements for both PTL and DG-A-FET-based circuits. As the inverters provide both complementary signals and external power for signal integrity, their dual-purpose use makes PTL extremely well-suited for use with DG-A-FETs in compact and efficient logic circuits. This paper therefore describes the A-PTL family and explores the A-PTL design space to propose and analyze three A-PTL structures with distinct trade-offs in terms of area, speed, and energy. These A-PTL variants are shown to provide propagation delay reduction of up to 47\% and energy savings of up to 88\% in comparison to the conventional CMOS-like logic family with the same A-FETs, leading to a 9x reduction in EDP and 20x reduction in area-energy-delay product (AEDP).\par

	\section{Background}
	
	In order to appreciate the potential benefits of using DG-A-FETs within the PTL family for large-scale computing systems, it is important to understand the potential and unique characteristics of these unconventional devices and of the PTL family. \par
	
	\subsection{Dual-Gate Ambipolar Carbon Nanotube Field-Effect Transistor}

    Though the high current density and tunable bandgap of low-dimensional materials such as CNTs have attracted significant attention, the presence of ambipolar transport has impeded attempts to use them to replace Si in unipolar FETs. However, this ambipolarity creates new opportunities for circuit design, as DG-A-FETs can be dynamically switched between \textit{n}-type and \textit{p}-type polarity.
    
    In the device of Fig. \ref{device}, the PG voltage determines the channel polarity while the CG voltage modulates current flow through the channel. This ability to switch the transistor between electron and hole conduction by the dual independent gate control allows the transistor to have more expressive power, reducing the number of devices required to perform logical operations \cite{Ben-Jamaa2011b}. In particular, the DG-A-FET of Fig. \ref{XNOR1}(a) natively provides a single-transistor XNOR operation and enables high-efficiency logical and neuromorphic computing systems \cite{Kenarangi2019d}. \par

    \subsection{Complementary Ambipolar Field-Effect Transistor Logic}

    When DG-A-FETs are used to implement logic circuits within a conventional complementary pull-up and pull-down logic structure, as in \cite{Ben-Jamaa2011b}, transmission gates (TGs) with complementary inputs are used in the pull-up and pull-down networks to prevent $\mathrm{V_{T}}$-drops \cite{Hu2017n}. For example, although a single DG-A-FET is able to implement the XNOR function, four DG-A-FETs are used to ensure signal integrity and fan-out as in Fig. \ref{XNOR1}(b). The requirement of the complementary transistor pair significantly increases the device count and hence the area cost and power consumption. Furthermore, since each transistor pair requires its input signal to be provided in a complementary form (both A and its complement are needed for A XNOR B), the overhead circuits and interconnects required to generate the complementary signal further reduce the benefits derived from utilizing transistor ambipolarity.\par
    
    Compared to conventional CMOS logic with unipolar transistors, DG-A-FETs within a conventional complementary pull-up and pull-down logic structure implement complex logic functions with higher logical expressivity. However, both the extra transistors used for ambipolar TG pairs and the inverters required for generating complementary signals decrease the efficiency of using A-FETs in the conventional CMOS logic architecture.  It is therefore worthwhile to explore other logic styles that are well-suited to logic with DG-A-FETs.\par

	\subsection{Pass Transistor Logic}
	
	PTL is a logic family designed to minimize the number of transistors required to implement any given logic function, thereby reducing area and power consumption. However, as there is no connection to a power supply, the energy for each PTL gate comes from its input, therefore necessitating the use of inverters to ensure signal integrity in cascaded circuits \cite{Zimmermann1997, Yano1996}. Furthermore, as traditional PTL uses only NMOS transistors, there is a threshold voltage drop whenever a high voltage signal is transmitted; an additional PMOS transistor is therefore frequently added to provide a full voltage swing using TGs in complementary PTL (CPL), which further necessitates inverters to provide each logical input signal in its complementary form. This heavy use of inverters and the need to use complementary transistors drastically increases the device count, delay, and energy consumption of PTL circuits, inhibiting the development of computing systems based entirely on PTL.\par
	
	PTL and CPL are therefore often used for specific logic functions within systems based primarily on the conventional complementary pull-up and pull-down logic structure. For example, multiplexing is particularly efficient with PTL, and PTL multiplexers are therefore frequently found within an otherwise conventional CMOS logic architecture. In that vein, previous work has proposed PTL and CPL multiplexers with DG-A-CNTFETs \cite{Jabeur2011} as well as a PTL full adder with ambipolar silicon nanowire FETs \cite{Zhang2014b}. These logic circuits, however, face the same challenges that have plagued other A-FETs and PTL: complementary input signals are required for functionality, requiring the heavy use of additional inverters. The area and energy costs of these inverters drastically decrease the efficiency, limiting the use of such circuits to particular logic functions within systems based primarily on the conventional complementary pull-up and pull-down logic structure. While previous DG-A-FET and PTL/CPL circuits are efficient for particular individual functions, novel approaches are required to enable a scalable logic family.

	\begin{figure}[t]
		\centering
		\includegraphics[width=0.35\textwidth]{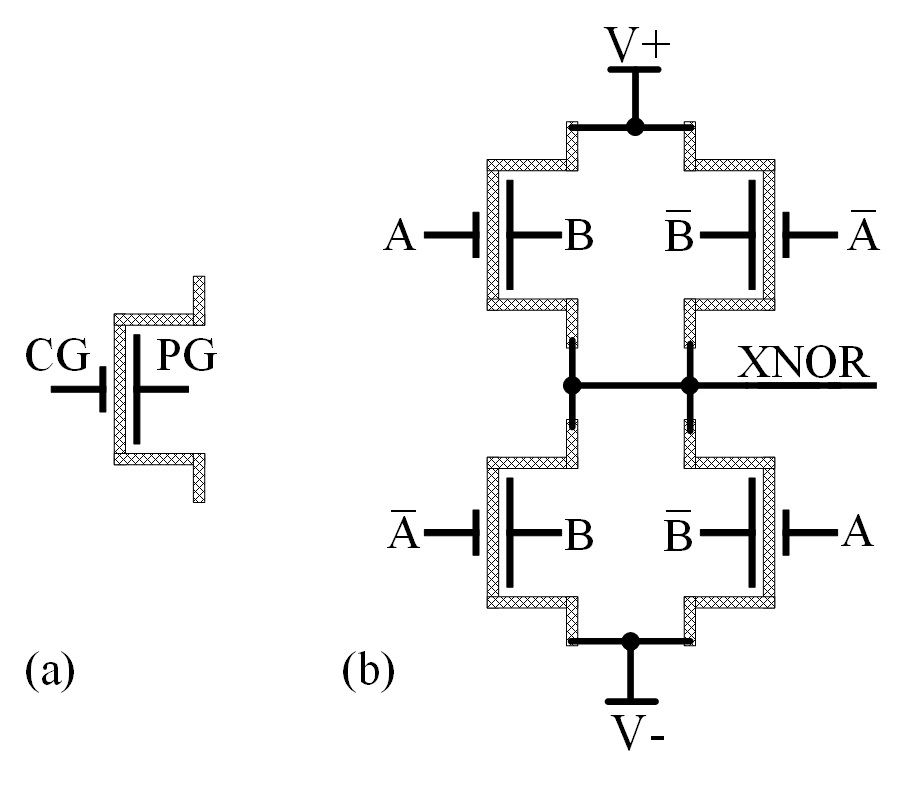}

		\vspace{-0.8em}
			\setcounter{figure}{1}
		\caption{Transistor-level schematic of XNOR gate with (a) single transistor, and (b) four transistors.}
		\vspace{-1.5em}
		\label{XNOR1}
	\end{figure}
	
	\section{Hybrid Ambipolar Pass Transistor Logic}
	
	The challenges facing both PTL and DG-A-FETs can be resolved by incorporating DG-A-FETs into a hybrid A-PTL family that can be scaled to large circuits and systems. In the proposed hybrid A-PTL family, the inverters required for complementary inputs and signal integrity simultaneously satisfy the needs of both DG-A-FETs and PTL, thereby amortizing the costs of these inverters across the benefits provided by both DG-A-FETs and PTL. Various hybrid A-PTL circuit styles are optimized for particular figures of merit, providing distinct advantages in large-scale circuits implemented entirely with hybrid A-PTL.

	\subsection{Basic Hybrid A-PTL Structure (Type I)}
	
    As illustrated in Fig. \ref{PTL_Diagram}, the core concept of the proposed hybrid A-PTL family is the alternation between PTL-based computational circuits and buffering inverters. The entire system is composed solely of DG-A-FETs, with CPL-based computations performed by ambipolar transmission gate cores (A-TGCs) followed by two CMOS-style DG-A-FET inverters. The A-TGC performs logical operations based on the input data, and is succeeded by two inverters that ensure signal integrity and provide the complementary output signals for cascading stages. These buffering inverters provide energy from the power supply (rather than from the input as in PTL), thereby boosting the fan-out and enabling large cascaded circuits without signal degradation. Furthermore, using DG-A-FETs to reduce the number of transistors in the PTL-style logic circuits reduces area, delay, and energy in comparison to PTL.\par

	An example hybrid A-PTL circuit is illustrated in the schematic of Fig. \ref{Sch_typeA}, which will henceforth be referred to as the Type I cascading style. This one-bit requires eight transistors in the A-TGC and an additional eight transistors in the four inverters, for a total of 16. In comparison to the 28 DG-A-FETs required for a one-bit full adder in the conventional CMOS architecture \cite{Hu2017n}, this represents a 43\% reduction in area. Furthermore, as demonstrated in the simulations of section IV, this hybrid A-PTL full adder structure exhibits significant reductions in delay and energy consumption.\par

	\begin{figure}[t]
		\vspace*{1em}
		\centering
		\includegraphics[width=0.48\textwidth]{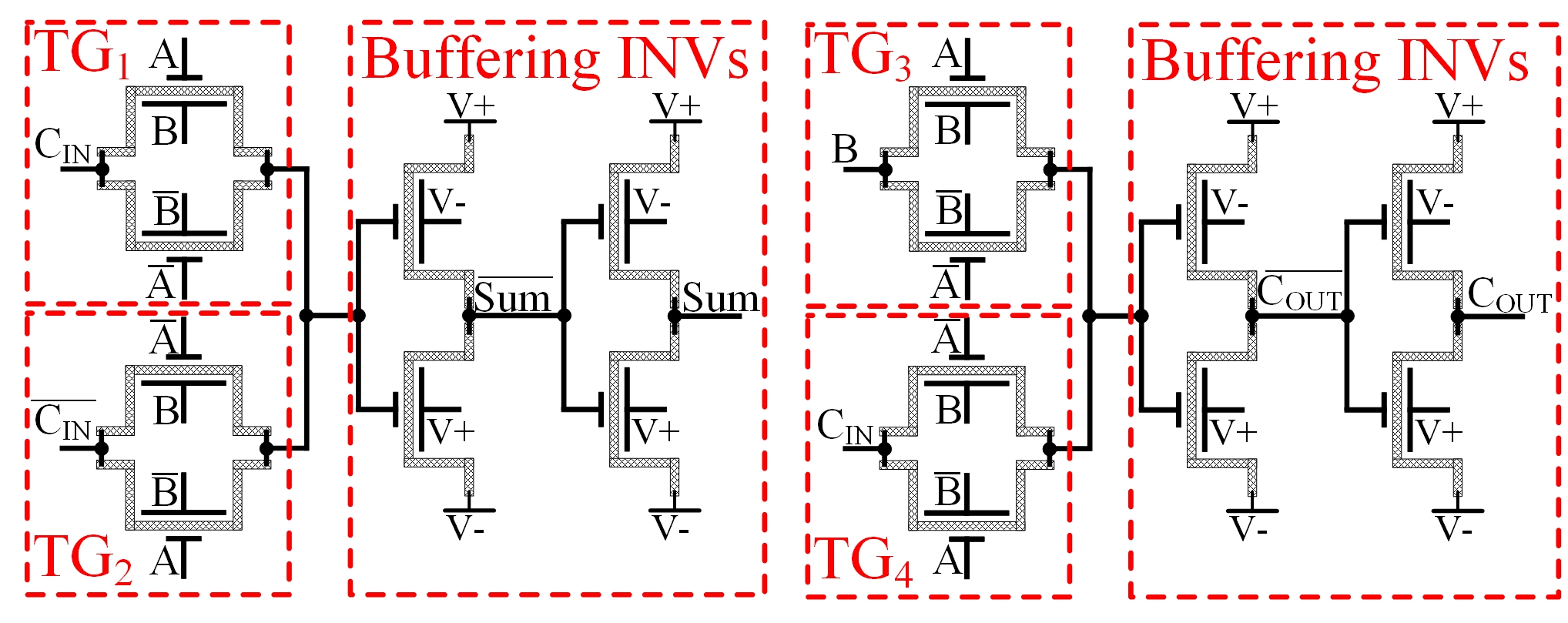}
		\setcounter{figure}{3}
		\caption{Transistor-level schematic of the Type I hybrid-A-PTL one-bit full adder.}
		\vspace{-0.5em}
		\label{Sch_typeA}
	\end{figure}

	\begin{figure}[t]
		\centering
		\includegraphics[trim= 120 400 160 240, clip, width=0.49\textwidth]{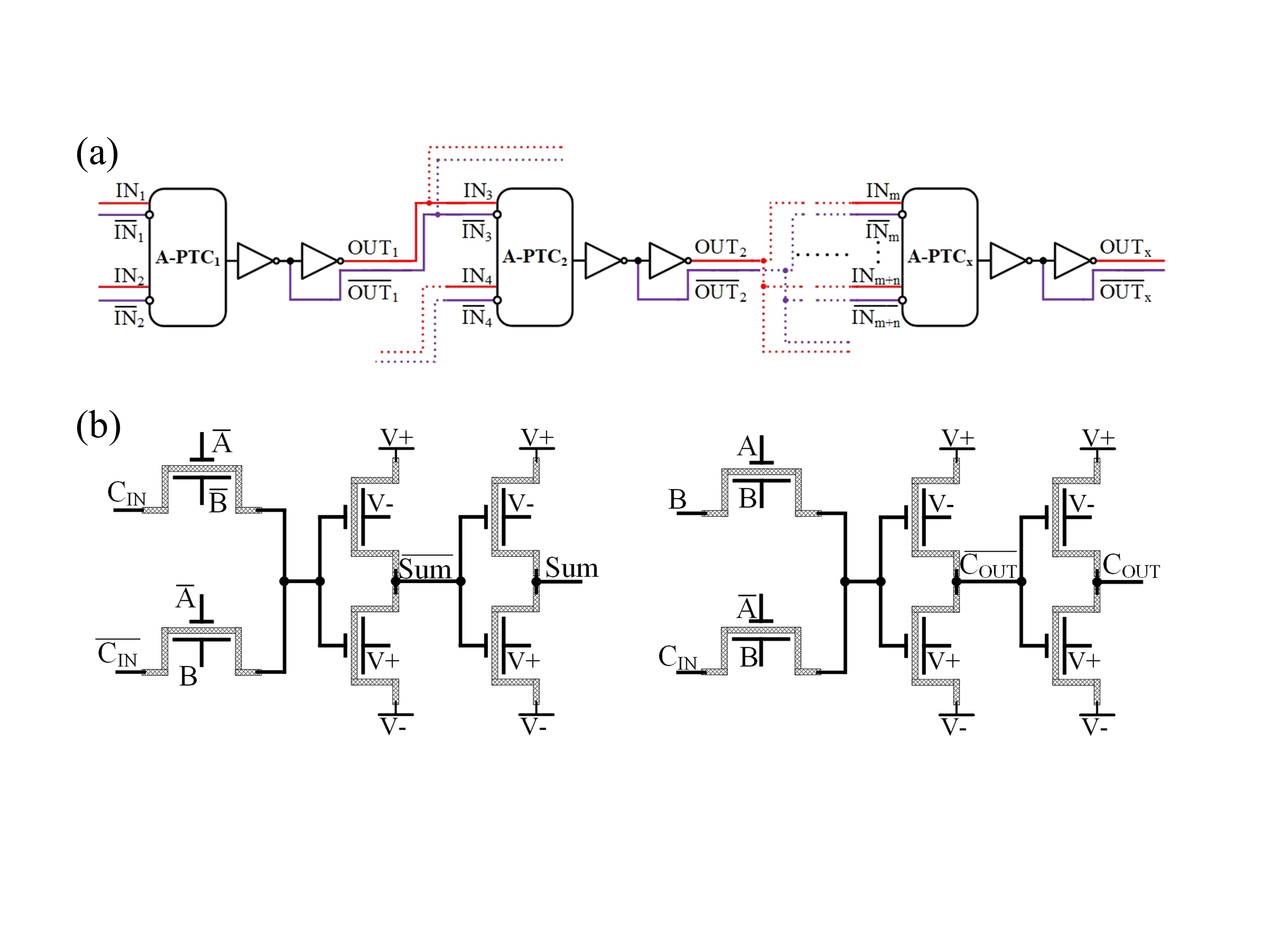}
		
		\caption{(a) Block diagram of the Type II hybrid A-PTL style. (b) Transistor-level schematic of a Type II hybrid-A-PTL one-bit full adder, showing A-TGCs replaced by A-PTCs.}
		\vspace{0.5em}
		\label{Sch_typeB}
	\end{figure}	
	
	\begin{figure}[t]
		\centering
		\includegraphics[trim= 80 300 650 240, clip, width=0.42\textwidth]{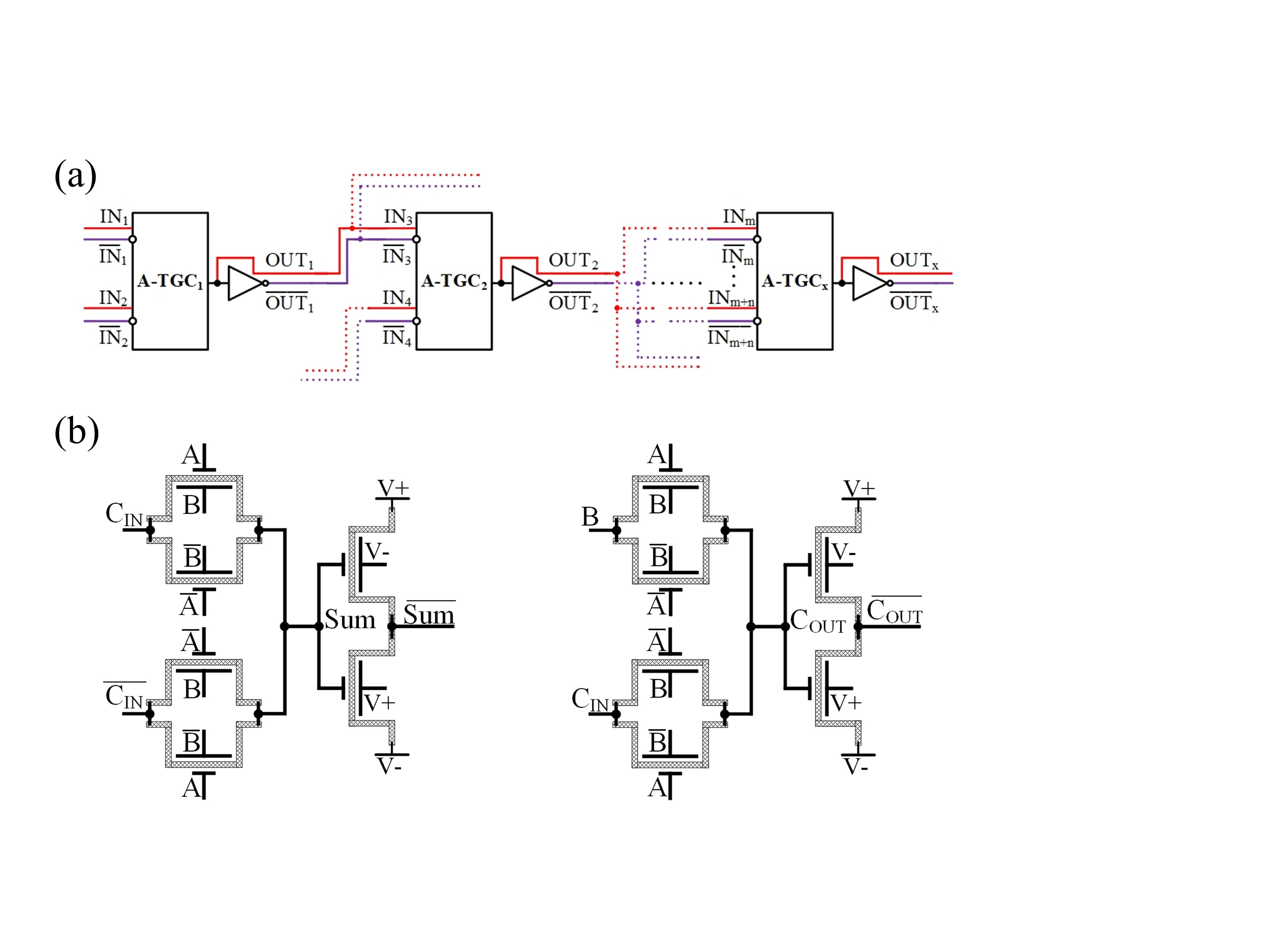}
		\caption{(a) Block diagram of the Type III hybrid A-PTL architecture showing the direct cascading of the non-inverted output without a buffering inverter. (b) Transistor-level schematic of a Type III hybrid-A-PTL one-bit full adder with only one inverter following each A-TGC.}
		\label{Sch_typeC}
	\end{figure}

\begin{table}[!t]
	\renewcommand{\arraystretch}{1.5}
	\caption{One-Bit Full Adder Area Comparison with Hybrid A-PTL}
	\label{DeviceCount_Table}
	\centering
	\begin{tabular}{c|c|c}
		\hline
		Architecture & Device Count & Estimated Area Reduction\\
		\hline
		CMOS-like 		& 28 & baseline\\
		Type I 			& 16 & 43\%	\\
		Type II			& 12 & 57\%	\\
		Type III 		& 12 & 57\%	\\
		\hline
	\end{tabular}
\end{table}

	\subsection{Alternative Hybrid A-PTL Structures (Types II \& III)}
	
	Beyond the proposed Type I hybrid A-PTL style, the transistor count can be minimized further by reducing complementarity in either the A-TGC or buffering inverters. These two alternative logic styles make the following modifications:
	\begin{itemize}
		\item Type II: one transistor is removed from each TG pair within the A-TGC.
		\item Type III: one buffering inverter is removed following each A-TGC.
    \end{itemize}
	Each of these modifications directly causes a significant decrease in the circuit area, while also drastically impacting energy consumption and delay.
	\par
	
	The Type II style is illustrated in Fig. \ref{Sch_typeB}, with each TG pair in the A-TGC replaced by an ambipolar pass transistor core (A-PTC) with a single pass transistor to reduce device count. As a single pass transistor cannot always provide a full voltage swing, the logical complexity of the Type II A-PTL is limited by the DG-A-FET threshold voltage drops from propagating a '1' ('0') when the DG-A-FET has an \textit{n}-type (\textit{p}-type) channel. As demonstrated in section IV, the area reduction in this Type II structure therefore comes at the cost of increased delay, limiting its overall efficiency. \par 
	
	The Type III structure shown in Fig. \ref{Sch_typeC} uses the same A-TGC structure as Type I, but one buffering inverter is removed. Therefore, the non-inverted output signals ($Sum$ and $C_\mathrm{OUT}$ in the full adder) drive cascaded stages directly, without passing through a buffering inverter. Like conventional CPL, this may prevent long chains of cascaded circuits because the energy of the entire CPL block comes from the initial input signals. Whereas the Type II style compromises signal integrity by having an asymmetric signal swing at the output of the A-PTC, the Type III style has reduced signal integrity due to the lack of isolation/buffering inverters between the input and output. As demonstrated in section IV, the removal of an inverter from the critical path gives the Type III style the lowest delays for circuits without high fan-out.\par
	
    Table \ref{DeviceCount_Table} summarizes the transistor count of one-bit full adders in each of the three styles. Compared to the CMOS-like one-bit full adder baseline \cite{Hu2017n}, the proposed Type I style reduces the area by 43\%, while Types II and III reduce the area by 57\%. The reduced circuit symmetry with both Types II and III decreases the output slew rate, increasing the delay. However, this is compensating to varying degrees by reductions in the parasitic capacitances and, for Type III, the removal of an inverter from the critical path.

	\begin{figure}[t]
		\vspace*{1em}
		\centering
		\includegraphics[trim= 150 320 360 360, clip, width=0.5\textwidth]{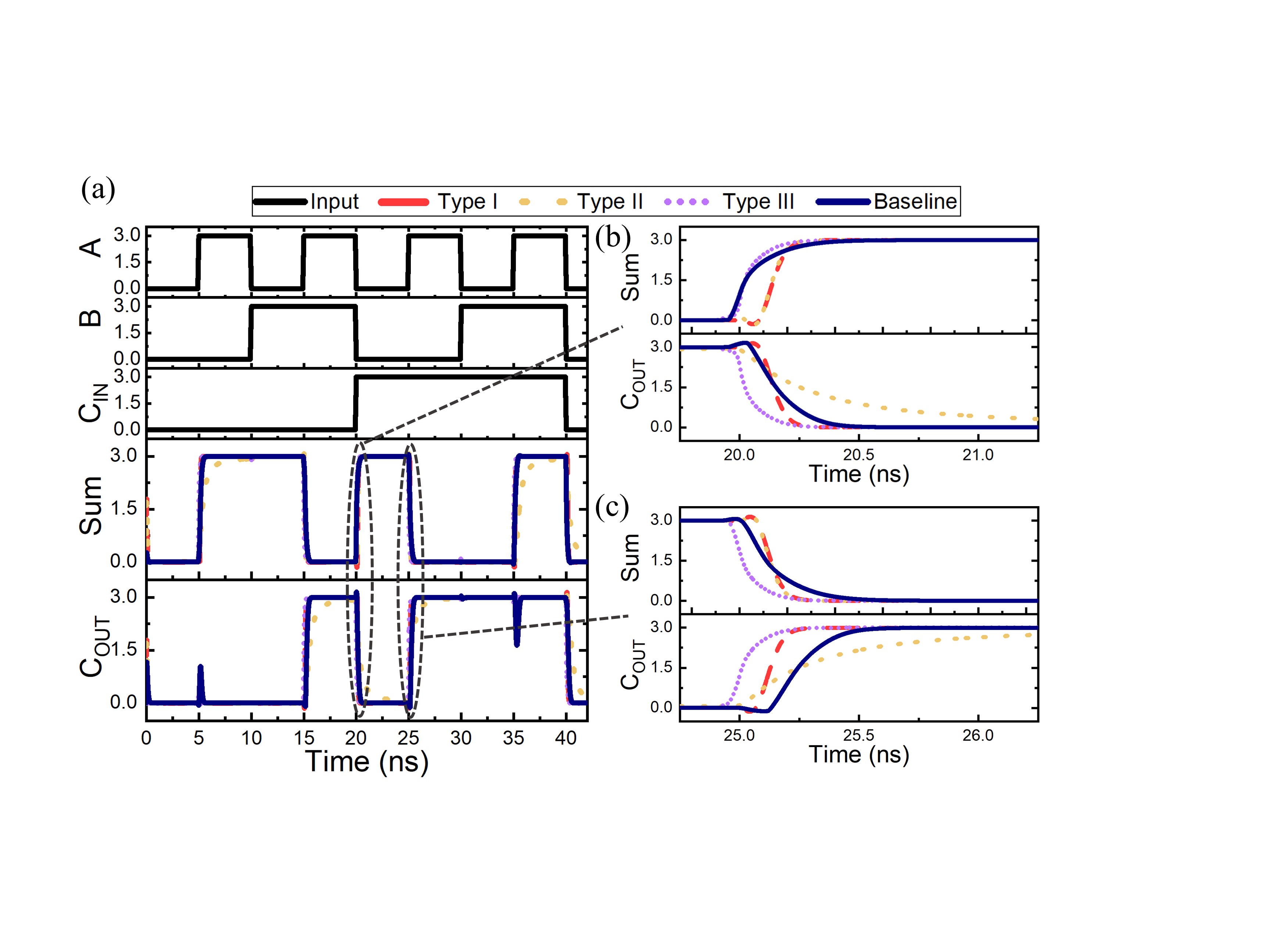}
		\vspace{-1.8em}
		\caption{Transient waveforms of the A-PTL one-bit full adders with CG cascading where (a) shows all input combinations while (b) and (c) show zoomed-in transitions of rising and falling edges.}
		\vspace{-0.5em}
		\label{Sim_TG}
	\end{figure}

	\begin{figure}[t]
		\centering
		\begin{subfigure}[t]{0.5\textwidth}
			\centering\raisebox{0em}{
				\includegraphics[width=0.5\textwidth]{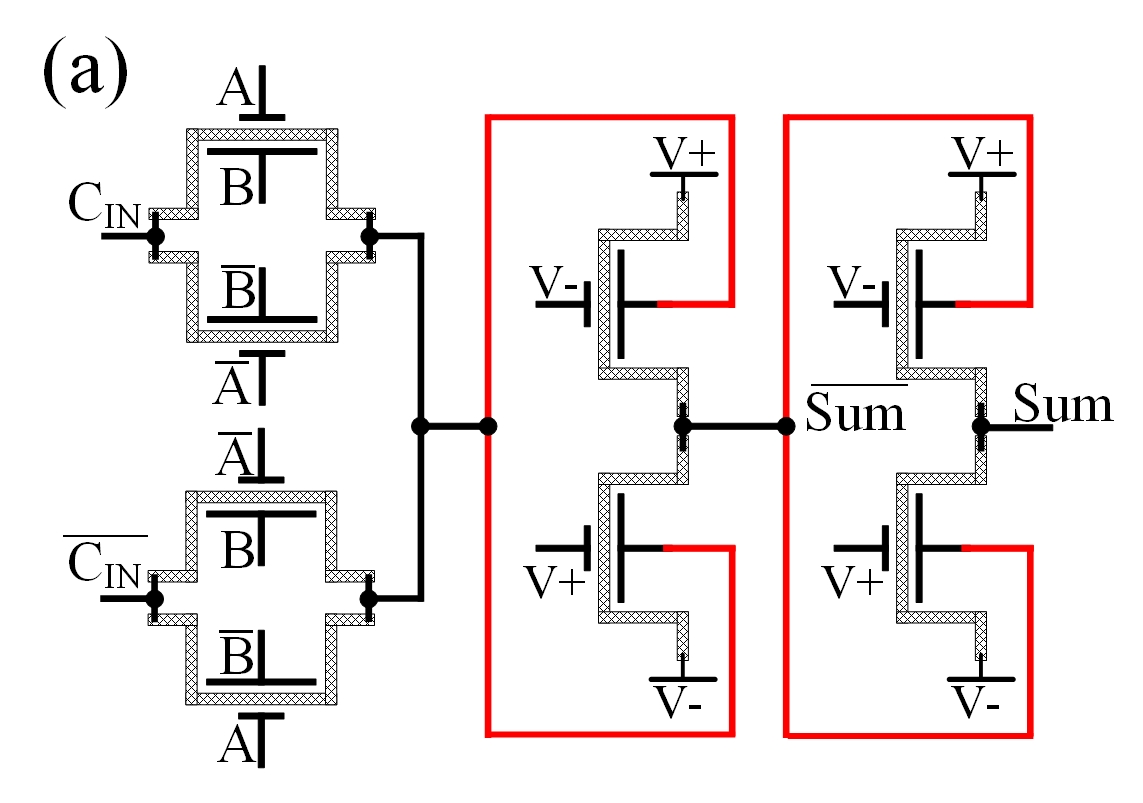}
			}
		\end{subfigure}\qquad

		\begin{subfigure}[t]{0.24\textwidth}
			\centering\raisebox{0.5em}{
				\includegraphics[width=1\textwidth]{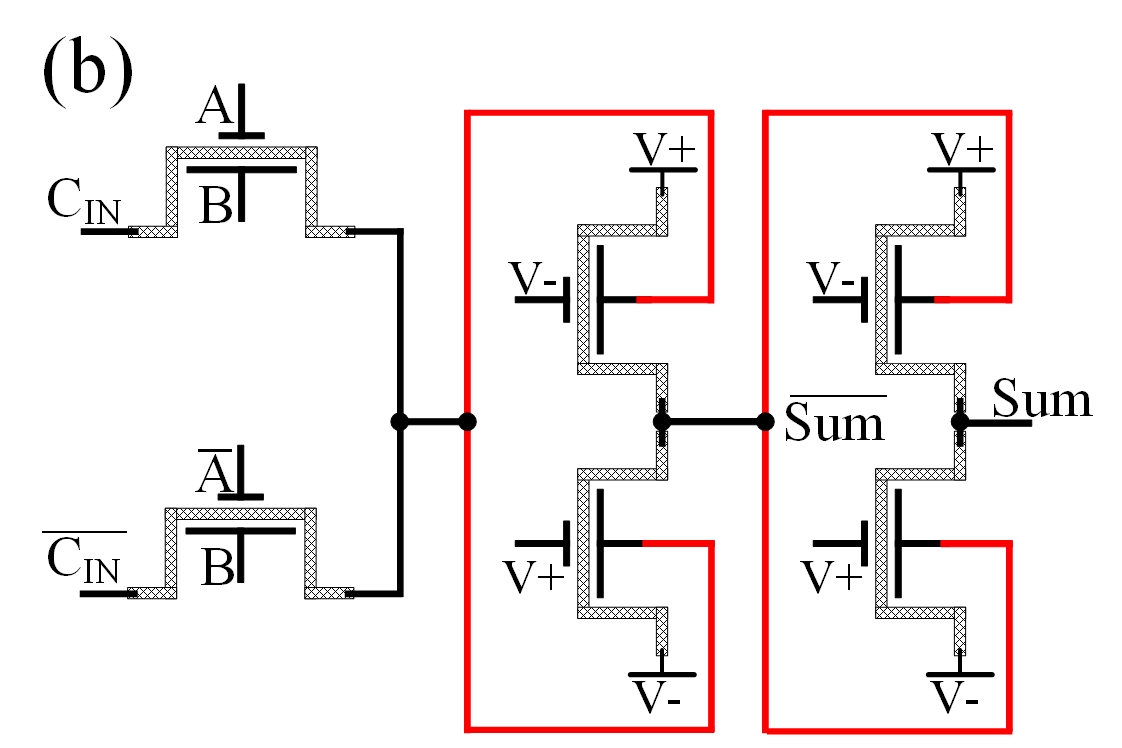}
			}
		\end{subfigure}\qquad
		\begin{subfigure}[t]{0.21\textwidth}
			\includegraphics[width=1\textwidth]{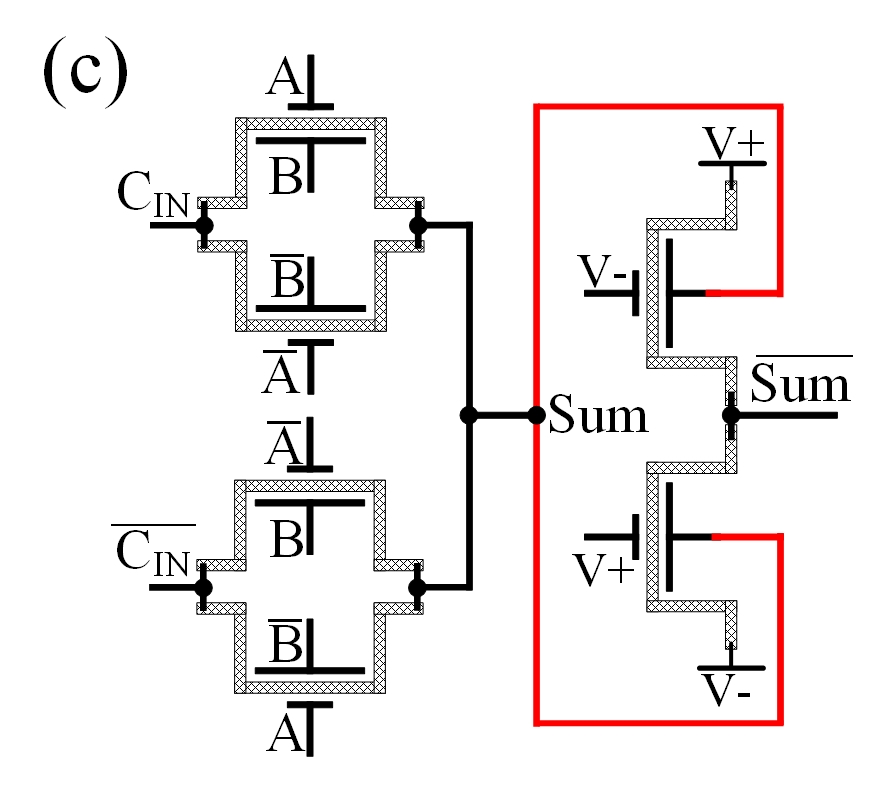}
		\end{subfigure}
		\caption{$Sum$ circuit schematic with PG cascading in (a) Type I, (b) Type II, and (c) Type III hybrid A-PTL styles.}
		\vspace{-1em}
		\label{Sch_BG}
	\end{figure}
	
	\begin{figure}[t]
		\vspace*{0em}
		\centering
		\includegraphics[trim= 200 320 360 300, clip, width=0.5\textwidth]{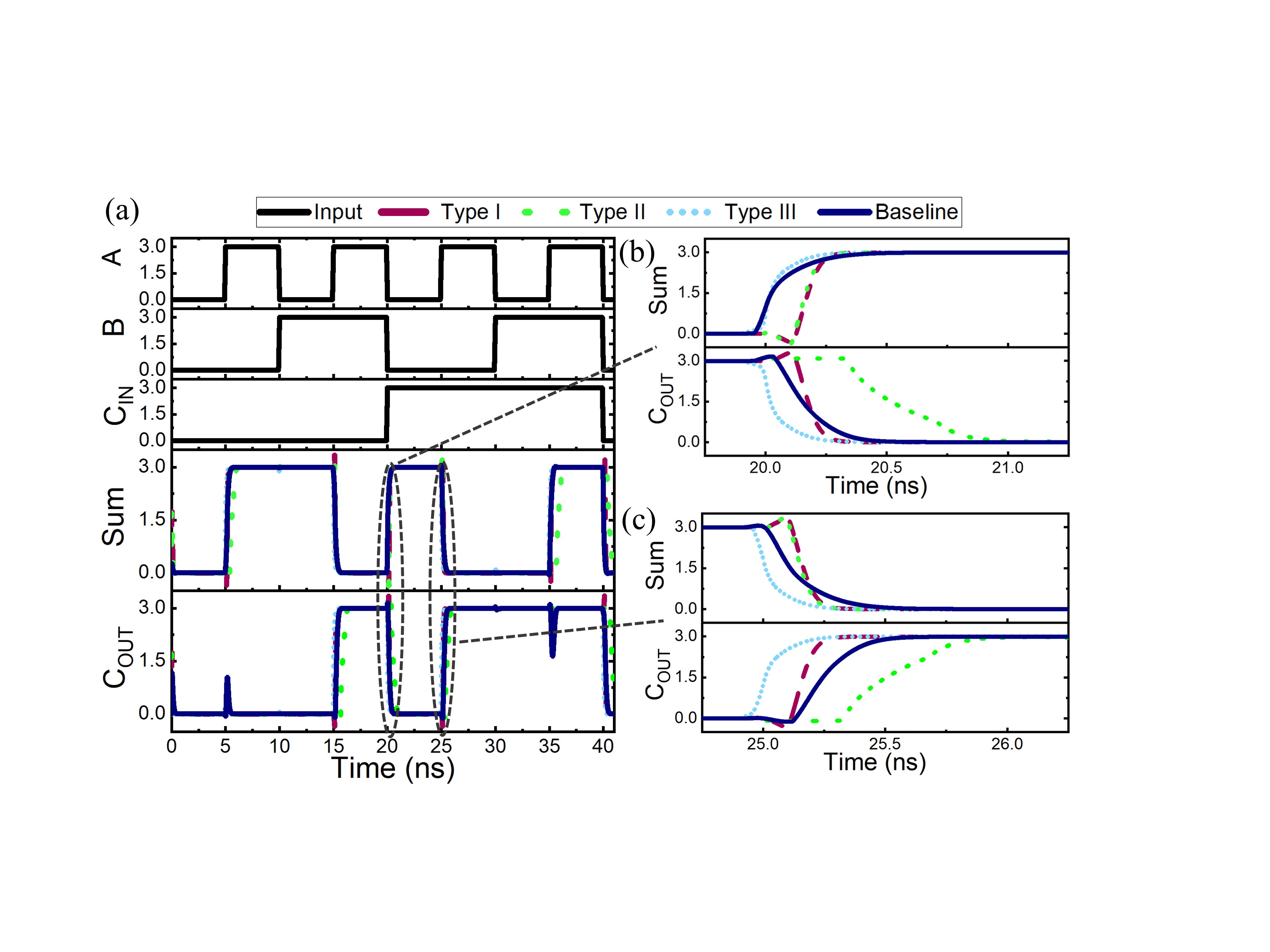}
		\vspace{-1.5em}
		\caption{Transient waveforms of the A-PTL one-bit full adders with PG cascading where (a) shows all input combinations while (b) and (c) show zoomed-in transitions of rising and falling edges.}
		\vspace{-1.5em}
		\label{Sim_BG}
	\end{figure}

	\begin{figure}[t]
		\centering
		\includegraphics[trim= 5 75 5 20, clip, width=0.5\textwidth]{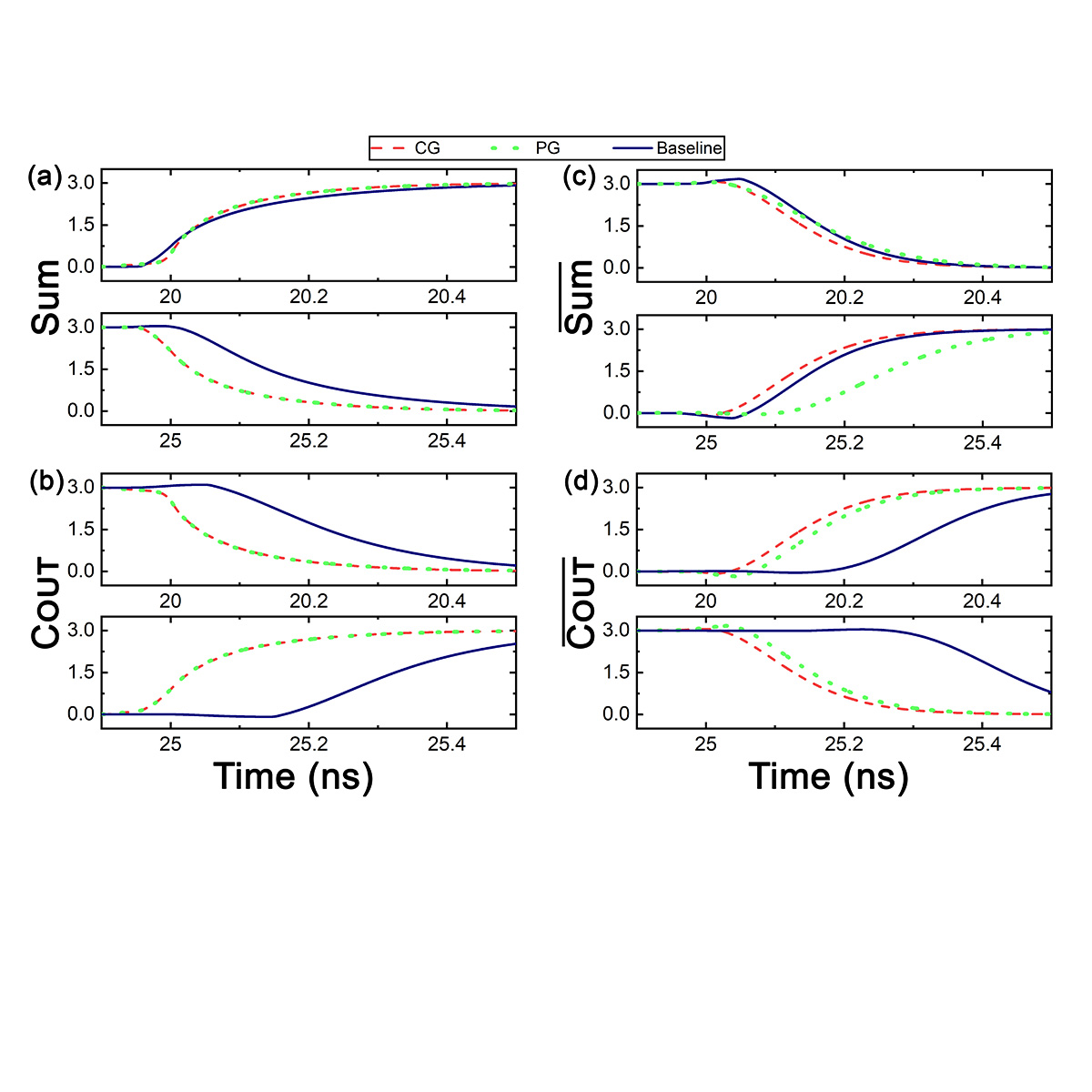}
		\caption{Transient switching waveforms of the Type III hybrid A-PTL one-bit full adder with CG cascading (dashed lines) and PG cascading (dotted lines) for: (a) $Sum$; (b) $C_\mathrm{OUT}$; (c) $\overline{Sum}$; (d) $\overline{C_\mathrm{OUT}}$.		}
		\vspace{-1.5em}
		\label{Sim_TGBG}
	\end{figure}

	\section{Computational Performance \& Efficiency}
	
	To evaluate the performance and efficiency of the proposed hybrid A-PTL styles, SPICE simulations have been performed to compare the various styles to the conventional complementary pull-up and pull-down structure \cite{Hu2017n}. While numerous ambipolarity has been demonstrated in numerous materials including silicon nanowires, $\mathrm{WSe_2}$, and graphene nanoribbons, the DG-A-CNTFET has been chosen for this analysis due its prior experimental demonstration and the availability of a SPICE-compatible device model \cite{Lin2004b,Lin2005,Hu2017a, Hu2017n}. SPICE simulations of one-bit and four-bit full adders demonstrate that the choice between CG and PG cascading engenders a trade-off between delay and energy consumption. Overall, the Type III style is shown to be superior, with delay reduced by a factor of two, energy reduced by a factor of eight, and AEDP reduced by a factor of 20.
	
	\subsection{Comparison of One-Bit Full Adders}
	
	Fig. \ref{Sim_TG} shows the transient simulation waveforms of one-bit full adder circuits in the three hybrid A-PTL styles. To provide a reference that enables apples-to-apples comparisons of the circuit structures, the conventional CMOS-like 28-transistor one-bit full adder of \cite{Hu2017n} servers as a baseline. As shown in Figs. \ref{Sch_typeA}-\ref{Sch_typeC}, the output of the A-TGC or A-PTC is fed to the CGs of the DG-A-FETs in the buffering inverters; the PGs of the DG-A-FETs in the inverters are connected to power supplies for DC biasing. Figs. \ref{Sim_TG}(b) and (c) zoom in on transitions of both the $Sum$ and $C_\mathrm{OUT}$ output signals. These simulations clearly reveal that the Type III style provides the shortest delay, significantly faster than the baseline. The Type I and II styles have varying delays in each case, though the Type II style is clearly quite slow for computing $C_\mathrm{OUT}$.\par

	\begin{figure*}[t]
		\centering
		\includegraphics[trim= 00 00 00 00, clip, width=0.95\textwidth]{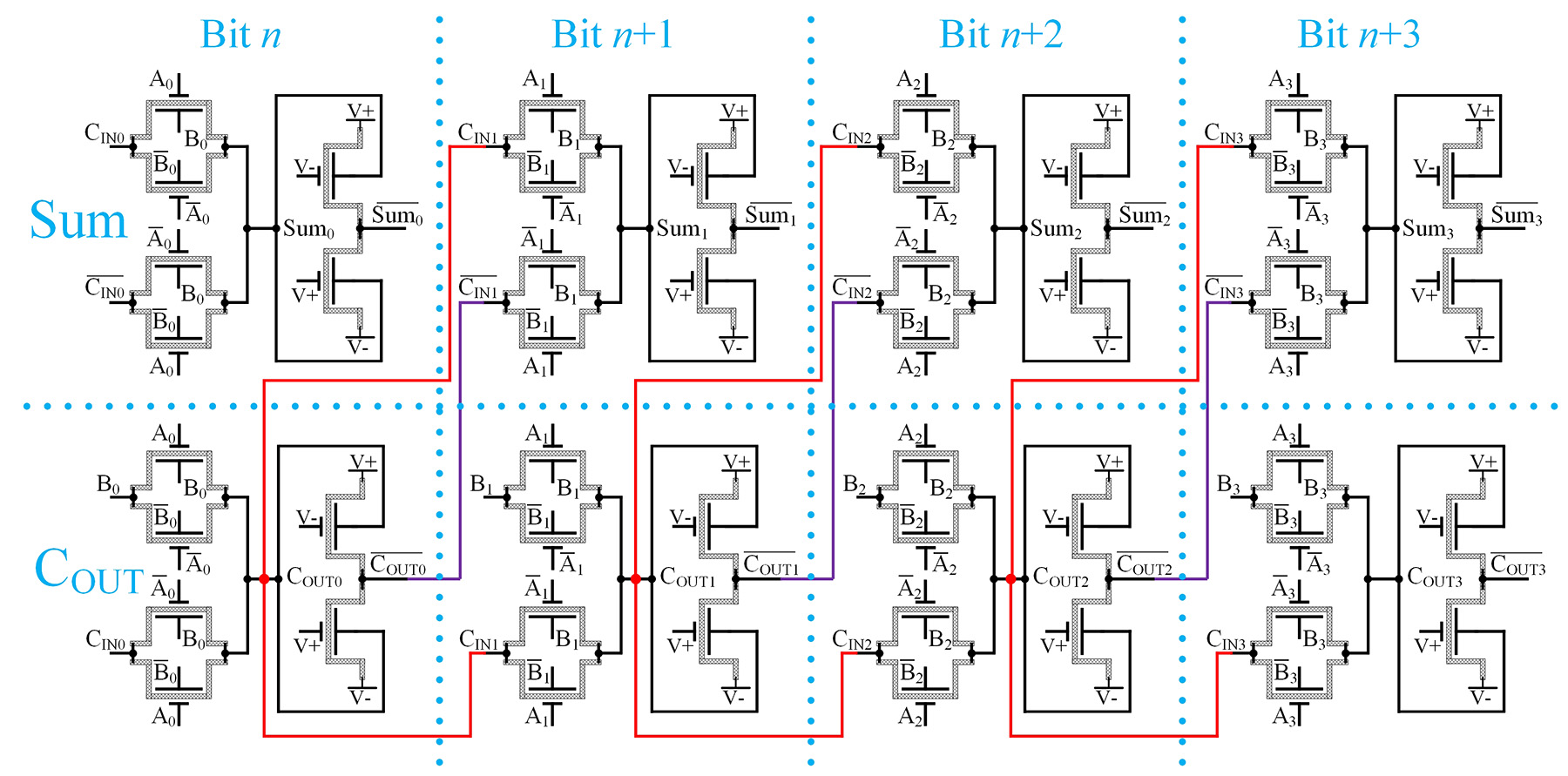}
		\caption{Transistor-level schematic of the hybrid A-PTL four-bit ripple-carry adder using the Type III style and PG cascading.}
		\label{Sch_Four}
	\end{figure*}
	
	\begin{figure}[t]
		\centering
		\includegraphics[trim= 7 52 11 37, clip, width=0.45\textwidth]{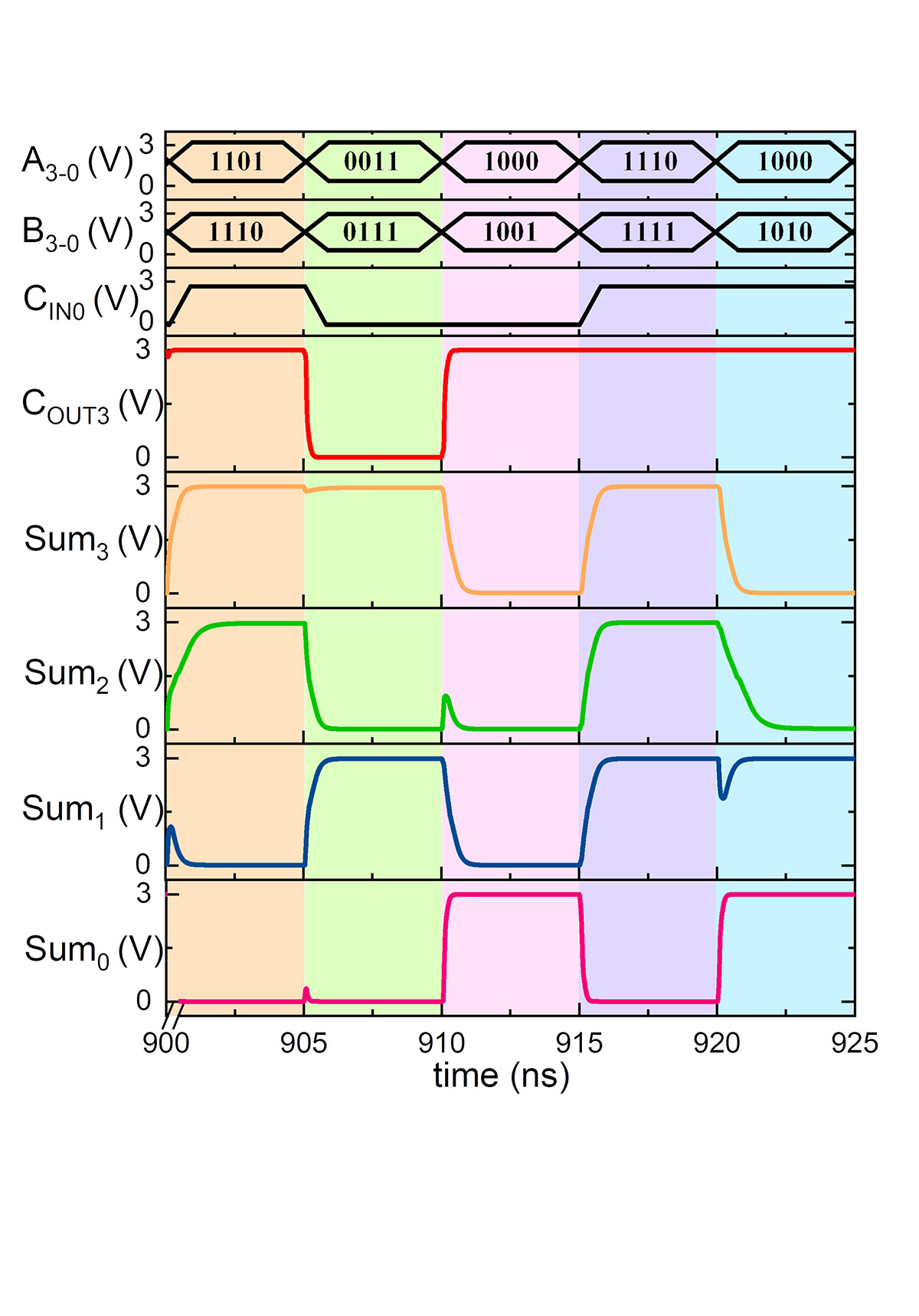}
		\vspace{-1.5em}
		\caption{Selected simulation waveforms of the Type III hybrid A-PTL four-bit ripple-carry adder with PG cascading.}
		\vspace{-1.5em}
		\label{Wave_four}
	\end{figure}

	\begin{figure*}[t]
		\centering
		\includegraphics[trim= 20 00 10 20, clip, width=0.95\textwidth]{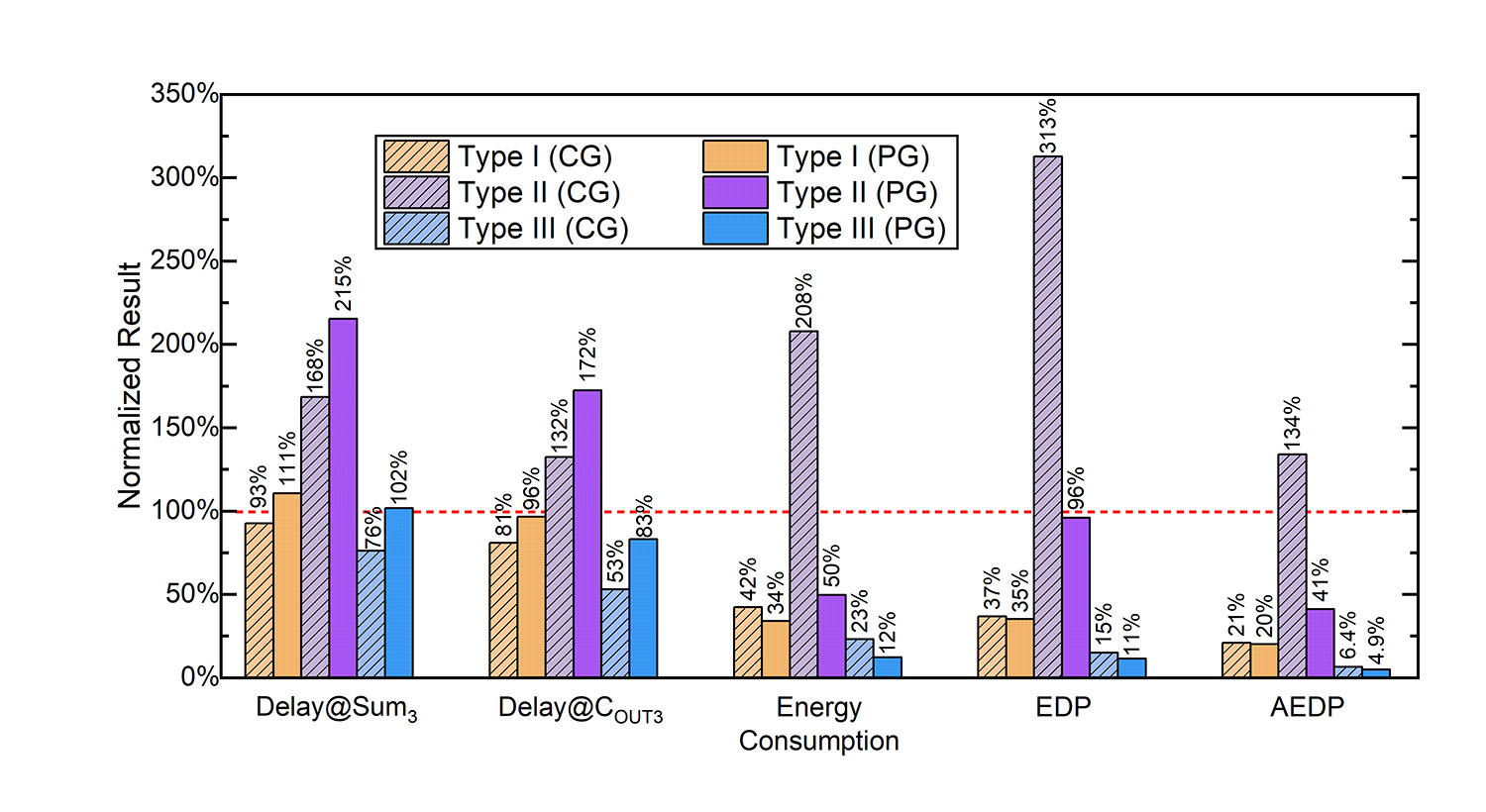}
		\vspace{-1.5em}
		\caption{Performance and efficiency comparison of the proposed hybrid A-PTL styles to the conventional CMOS-style baseline. These results are based on simulations of the four cascaded full adders.}
		\vspace{-1.5em}
		\label{Sim_all}
	\end{figure*}

    \subsection{Control Gate vs. Polarity Gate Cascading}
	In all of the circuits that have thus far been discussed, the output of the A-TGC or A-PTC has been fed to the CGs of the DG-A-FETs within the inverters. With this CG cascading, the polarity of each DG-A-FET within the inverters is constant. Alternatively, this signal could be used to drive the PGs within the inverter DG-A-FETs, as shown in Fig. \ref{Sch_BG}, thus broadening the design space. In the case of PG cascading, the the polarity of the DG-A-FETs within the inverters is modulated by the outputs of the A-TGC or A-PTC while the CG input remains constant.\par
	
	Fig. \ref{Sim_BG}(a) shows the transient simulation waveforms of one-bit full adder circuits with PG cascading in the three hybrid A-PTL styles, with a focus on the switching at selected edges in Figs. \ref{Sim_BG}(b) and (c). Similar to the simulations with CG cascading in Fig. \ref{Sim_TG}, these simulations show that the Type III style exhibits the shortest delay, while the relative delays of the Type I and II styles vary.\par
	
	To directly compare CG and PG cascading, the transient waveforms with both CG and PG cascading are plotted together for the Type III hybrid A-PTL full adder in Fig. \ref{Sim_TGBG}. As the non-inverted $Sum$ and $C_\mathrm{OUT}$ signals have not passed through an inverter, there is no difference between CG and PG cascading.  For the inverted output signals, this comparison clearly shows that CG cascading provides shorter delay than PG cascading due to an earlier transition. However, the PG cascading has a higher output slew rate, leading to lower energy consumption.\par
	
    \subsection{Medium-Scale Circuit Analysis}
    
    To enable analysis more predictive of a large-scale system, four-bit hybrid A-PTL ripple-carry adders have been simulated with all three styles and both cascading approaches, for a total of six distinct circuits. As an example, the four-bit hybrid A-PTL ripple-carry adder using the Type III style with PG cascading is shown in Fig. \ref{Sch_Four}. To broadly characterize the behavior across input combinations, a sequence of 1,000 operations was randomly generated and used as input signals.\par
    
    A selection of the transient simulation waveforms is shown in Fig. \ref{Wave_four}. This four-bit full adder circuit has an additional carry-in bit such that it is composed of four one-bit full adders rather than three one-bit full adders and a half adder. As can be seen in the figure, the missing inverter from the Type III style leads to a propagation delay dependent on the input combination.\par

    The performance and efficiency of all three hybrid A-PTL styles and both cascading approaches are analyzed in Fig. \ref{Sim_all} based on the results for the four-bit full adder circuits. These results are compared to the conventional CMOS-style DG-A-FET circuit as a baseline in terms of average delay, energy consumption, EDP, and AEDP across the 1,000 input combinations. As can clearly be observed in the figure, the Type III hybrid A-PTL style is superior to the conventional baseline structure for all metrics.\par
    
    The asymmetry between the CG and PG of the DG-A-FET devices leads to a trade-off between speed and energy consumption. For all three hybrid A-PTL styles, CG cascading is always faster than PG cascading; however, the circuits with PG cascading consume less energy and therefore result in a lower EDP and AEDP. Therefore, CG cascading is preferred when speed is critical, while PG cascading is more suitable to low-power system design. \par
    
    While the delay of the Type I and III styles can be significantly less than the baseline, the type II style is significantly slower. The Type I and III styles also provide enormous reductions in energy consumption, with the Type III style providing an 8x reduction when using PG cascading. The energy consumption of the Type II style strongly depends on the cascading approach, with PG cascading producing an energy consumption 4x less than CG cascading. The lack of complementary transistors in the Type II A-PTC results in lower signal integrity than with the A-TGC used in Types I and III, leading to increased EDP and AEDP for Type II.
    
    Hybrid A-PTL drastically improves the overall computing efficiency, with the Type III style with PG cascading providing a 9x improvement in EDP and a 20x improvement in AEDP. These figures of merit leverage the slightly improved delay and massively improved energy consumption, with the AEDP further incorporating the 57\% reduction in area (Table \ref{DeviceCount_Table}). While the Type I style is also promising, the reduced transistor count and simplified circuit structure provide the Type III style the greatest potential for compact and efficient circuits.\par

	\section{Conclusions}
	
	DG-A-FETs are naturally compatible with PTL, and the proposed hybrid A-PTL family leverages the advantages of both DG-A-FETs and PTL by applying their required inverters in a manner that efficiently amortizes their costs. This A-PTL family enables device count to be reduced by half, and increases energy-efficiency by an order of magnitude. Comparisons among the three hybrid A-PTL styles proposed in this work indicate that the Type III style is superior, thanks to the complementarity of the A-TGC and the removal of the second inverter. Furthermore, CG cascading is shown to be faster than PG cascading, while PG cascading is shown to be more energy-efficient than CG cascading. By comparing a medium-scale hybrid A-PTL circuit to one realized with DG-A-FETs in the conventional CMOS-like logic style, the hybrid A-PTL system is shown to provide up to 47\% decrease in delay, 57\% reduction in area, 88\% reduction in energy consumption, 9x reduction in EDP, and 20x reduction in AEDP. These results therefore greatly advance the already-promising prospects for efficient computing systems with ambipolar transistors.\par

	\bibliographystyle{myIEEEtran_X}
	\bibliography{TCASII_APTL,TCASII_APTL_patent}

\begin{thebibliography}{10}
\providecommand{\url}[1]{#1}
\csname url@samestyle\endcsname
\providecommand{\newblock}{\relax}
\providecommand{\bibinfo}[2]{#2}
\providecommand{\BIBentrySTDinterwordspacing}{\spaceskip=0pt\relax}
\providecommand{\BIBentryALTinterwordstretchfactor}{4}
\providecommand{\BIBentryALTinterwordspacing}{\spaceskip=\fontdimen2\font plus
\BIBentryALTinterwordstretchfactor\fontdimen3\font minus
  \fontdimen4\font\relax}
\providecommand{\BIBforeignlanguage}[2]{{%
\expandafter\ifx\csname l@#1\endcsname\relax
\typeout{** WARNING: IEEEtran.bst: No hyphenation pattern has been}%
\typeout{** loaded for the language `#1'. Using the pattern for}%
\typeout{** the default language instead.}%
\else
\language=\csname l@#1\endcsname
\fi
#2}}
\providecommand{\BIBdecl}{\relax}
\BIBdecl

\bibitem{Lin2004b}
Y.-M. Lin, J.~Appenzeller, P.~Avouris, and Watson, ``{Novel carbon nanotube FET
  design with tunable polarity},'' in \emph{IEDM}, 2004, pp. 687--690.

\bibitem{Lin2005}
\BIBentryALTinterwordspacing
Y.-M. Lin, J.~Appenzeller, J.~Knoch, and P.~Avouris, ``{High-performance carbon
  nanotube field-effect transistor with tunable polarities},'' \emph{IEEE
  Trans. Nanotechnology}, vol.~4, no.~5, pp. 481--489, 2005.
\BIBentrySTDinterwordspacing

\bibitem{Hu2017a}
\BIBentryALTinterwordspacing
X.~Hu and J.~S. Friedman, ``{Closed-form model for dual-gate ambipolar CNTFET
  circuit design},'' in \emph{2017 IEEE International Symposium on Circuits and
  Systems (ISCAS)}.\hskip 1em plus 0.5em minus 0.4em\relax IEEE, may 2017.
\BIBentrySTDinterwordspacing

\bibitem{Hu2017n}
\BIBentryALTinterwordspacing
X.~Hu and J.~S. Friedman, ``{Transient model with interchangeability for
  dual-gate ambipolar CNTFET logic design},'' in \emph{2017 IEEE/ACM
  International Symposium on Nanoscale Architectures (NANOARCH)}, vol.~5.\hskip
  1em plus 0.5em minus 0.4em\relax IEEE, jul 2017, pp. 61--66.
\BIBentrySTDinterwordspacing

\bibitem{Pan2018}
C.~Pan, Y.~Fu, J.~Wang, J.~Zeng, G.~Su, M.~Long, E.~Liu, C.~Wang, A.~Gao,
  M.~Wang, Y.~Wang, Z.~Wang, S.~J. Liang, R.~Huang, and F.~Miao, ``{Analog
  Circuit Applications Based on Ambipolar Graphene/MoTe2 Vertical
  Transistors},'' \emph{Advanced Electronic Materials}, vol.~4, no.~3, pp.
  1--7, 2018.

\bibitem{Bucella2016}
S.~G. Bucella, J.~M. Salazar-Rios, V.~Derenskyi, M.~Fritsch, U.~Scherf, M.~A.
  Loi, and M.~Caironi, ``{Inkjet Printed Single-Walled Carbon Nanotube Based
  Ambipolar and Unipolar Transistors for High-Performance Complementary Logic
  Circuits},'' \emph{Advanced Electronic Materials}, vol.~2, no.~6, pp. 1--6,
  2016.

\bibitem{Zhang2013c}
\BIBentryALTinterwordspacing
J.~Zhang, P.-E. Gaillardon, and G.~{De Micheli}, ``{Dual-threshold-voltage
  configurable circuits with three-independent-gate silicon nanowire FETs},''
  in \emph{2013 IEEE International Symposium on Circuits and Systems
  (ISCAS2013)}.\hskip 1em plus 0.5em minus 0.4em\relax IEEE, may 2013, pp.
  2111--2114.
\BIBentrySTDinterwordspacing

\bibitem{Zhang2014}
J.~Zhang, M.~D. Marchi, D.~Sacchetto, P.-E. Gaillardon, Y.~Leblebici, and G.~D.
  Micheli, ``{Polarity-controllable Silicon nanowire transistors with dual
  threshold voltages},'' \emph{IEEE Trans. Electron Devices}, vol.~61, no.~11,
  pp. 3654--3660, 2014.

\bibitem{Hu2019BookChapter}
\BIBentryALTinterwordspacing
X.~Hu, W.~H. Brigner, and {J. S. Friedman}, ``{CNT and SiNW modeling for
  dual-gate ambipolar logic circuit design},'' in \emph{Functionality-Enhanced
  Devices An alternative to Moore's Law}, P.-E. Gaillardon, Ed.\hskip 1em plus
  0.5em minus 0.4em\relax Institution of Engineering and Technology, nov 2018,
  ch.~8, pp. 151--188.
\BIBentrySTDinterwordspacing

\bibitem{Yin2017b}
\BIBentryALTinterwordspacing
L.~Yin, Z.~Wang, F.~Wang, K.~Xu, R.~Cheng, Y.~Wen, J.~Li, and J.~He,
  ``{Ferroelectric-induced carrier modulation for ambipolar transition metal
  dichalcogenide transistors},'' \emph{Applied Physics Letters}, vol. 110,
  no.~12, p. 123106, 2017.
\BIBentrySTDinterwordspacing

\bibitem{Zhang2012}
Y.~Zhang, J.~Ye, Y.~Matsuhashi, and Y.~Iwasa, ``{Ambipolar MoS 2 thin flake
  transistors},'' \emph{Nano Letters}, vol.~12, no.~3, pp. 1136--1140, 2012.

\bibitem{Kim2017b}
\BIBentryALTinterwordspacing
B.~Kim, M.~L. Geier, M.~C. Hersam, and A.~Dodabalapur, ``{Inkjet printed
  circuits based on ambipolar and p-Type carbon nanotube thin-film
  transistors},'' \emph{Scientific Reports}, vol.~7, pp. 1--8, 2017.
\BIBentrySTDinterwordspacing

\bibitem{Ortiz2018}
\BIBentryALTinterwordspacing
D.~N. Ortiz, I.~Ramos, N.~J. Pinto, M.~Q. Zhao, V.~Kumar, and A.~T. Johnson,
  ``{Ambipolar transport in CVD grown MoSe2 monolayer using an ionic liquid gel
  gate dielectric},'' \emph{AIP Advances}, vol.~8, no.~3, 2018.
\BIBentrySTDinterwordspacing

\bibitem{Tian2010}
J.~F. Tian, L.~A. Jauregui, G.~Lopez, H.~Cao, and Y.~P. Chen, ``{Ambipolar
  graphene field effect transistors by local metal side gates},'' \emph{Applied
  Physics Letters}, vol.~96, no.~26, pp. 94--97, 2010.

\bibitem{Yu2018}
M.~Yu, H.~Wan, L.~Cai, J.~Miao, S.~Zhang, and C.~Wang, ``{Fully Printed
  Flexible Dual-Gate Carbon Nanotube Thin-Film Transistors with Tunable
  Ambipolar Characteristics for Complementary Logic Circuits},'' \emph{ACS
  Nano}, vol.~12, no.~11, pp. 11\,572--11\,578, 2018.

\bibitem{Baeg2017}
\BIBentryALTinterwordspacing
K.~J. Baeg, H.~J. Jeong, S.~Y. Jeong, J.~T. Han, and G.~W. Lee, ``{Enhanced
  ambipolar charge transport in staggered carbon nanotube field-effect
  transistors for printed complementary-like circuits},'' \emph{Current Applied
  Physics}, vol.~17, no.~4, pp. 541--547, 2017.
\BIBentrySTDinterwordspacing

\bibitem{Ben-Jamaa2011b}
\BIBentryALTinterwordspacing
M.~H. Ben-Jamaa, K.~Mohanram, and G.~{De Micheli}, ``{An efficient gate library
  for ambipolar CNTFET logic},'' \emph{IEEE Transactions on Computer-Aided
  Design of Integrated Circuits and Systems}, vol.~30, no.~2, pp. 242--255, feb
  2011.
\BIBentrySTDinterwordspacing

\bibitem{Resta2016}
\BIBentryALTinterwordspacing
G.~V. Resta, S.~Sutar, Y.~Balaji, D.~Lin, P.~Raghavan, I.~Radu, F.~Catthoor,
  A.~Thean, P.-E. Gaillardon, and G.~de~Micheli, ``{Polarity control in WSe2
  double-gate transistors},'' \emph{Scientific Reports}, vol.~6, no.~1, p.
  29448, sep 2016.
\BIBentrySTDinterwordspacing

\bibitem{Yano1996}
\BIBentryALTinterwordspacing
K.~Yano, Y.~Sasaki, K.~Rikino, and K.~Seki, ``{Top-down pass-transistor logic
  design},'' \emph{IEEE Journal of Solid-State Circuits}, vol.~31, no.~6, pp.
  792--803, jun 1996.
\BIBentrySTDinterwordspacing

\bibitem{Radhakrishnan1985}
\BIBentryALTinterwordspacing
D.~Radhakrishnan, S.~Whitaker, and G.~Maki, ``{Formal design procedures for
  pass transistor switching circuits},'' \emph{IEEE Journal of Solid-State
  Circuits}, vol.~20, no.~2, pp. 531--536, apr 1985.
\BIBentrySTDinterwordspacing

\bibitem{Yano1990}
K.~Yano, A.~Shimizu, T.~Nishida, M.~Saito, and K.~Shimohigashi, ``{A 3.8-ns
  CMOS 16 X 16-b Multiplier Using Complementary Pass-Transistor Logic},''
  \emph{IEEE Journal of Solid-State Circuits}, vol.~25, no.~2, pp. 388--395,
  1990.

\bibitem{VonNeumann1993b}
\BIBentryALTinterwordspacing
J.~von Neumann, ``{First draft of a report on the EDVAC},'' Tech. Rep., 1945.
\BIBentrySTDinterwordspacing

\bibitem{Kenarangi2019d}
\BIBentryALTinterwordspacing
F.~Kenarangi, X.~Hu, Y.~Liu, J.~A.~C. Incorvia, J.~S. Friedman, and
  I.~Partin-Vaisband, ``{Exploiting dual-gate ambipolar CNFETs for scalable
  machine learning classification},'' \emph{Scientific Reports}, vol.~10,
  no.~1, p. 5735, dec 2020.
\BIBentrySTDinterwordspacing

\bibitem{Zimmermann1997}
R.~Zimmermann and W.~Fichtner, ``{Low-power logic styles: CMOS versus
  pass-transistor logic},'' \emph{IEEE Journal of Solid-State Circuits},
  vol.~32, no.~7, pp. 1079--1090, 1997.

\bibitem{Jabeur2011}
\BIBentryALTinterwordspacing
K.~Jabeur, I.~O'Connor, N.~Yakymets, and S.~{Le Beux}, ``{High performance 4:1
  multiplexer with ambipolar double-gate FETs},'' in \emph{IEEE ICECS},
  no.~c.\hskip 1em plus 0.5em minus 0.4em\relax IEEE, dec 2011, pp. 677--680.
\BIBentrySTDinterwordspacing

\bibitem{Zhang2014b}
J.~Zhang, X.~Tang, P.-E. Gaillardon, and G.~{De Micheli}, ``{Configurable
  circuits featuring dual-threshold-voltage design with three-independent-gate
  Silicon nanowire FETs},'' \emph{IEEE Trans. Circuits {\&} Systems I},
  vol.~61, no.~10, pp. 2851--2861, 2014.

\end{thebibliography}

\begin{IEEEbiography}[{\includegraphics[width=1in,height=1.25in,clip,keepaspectratio]{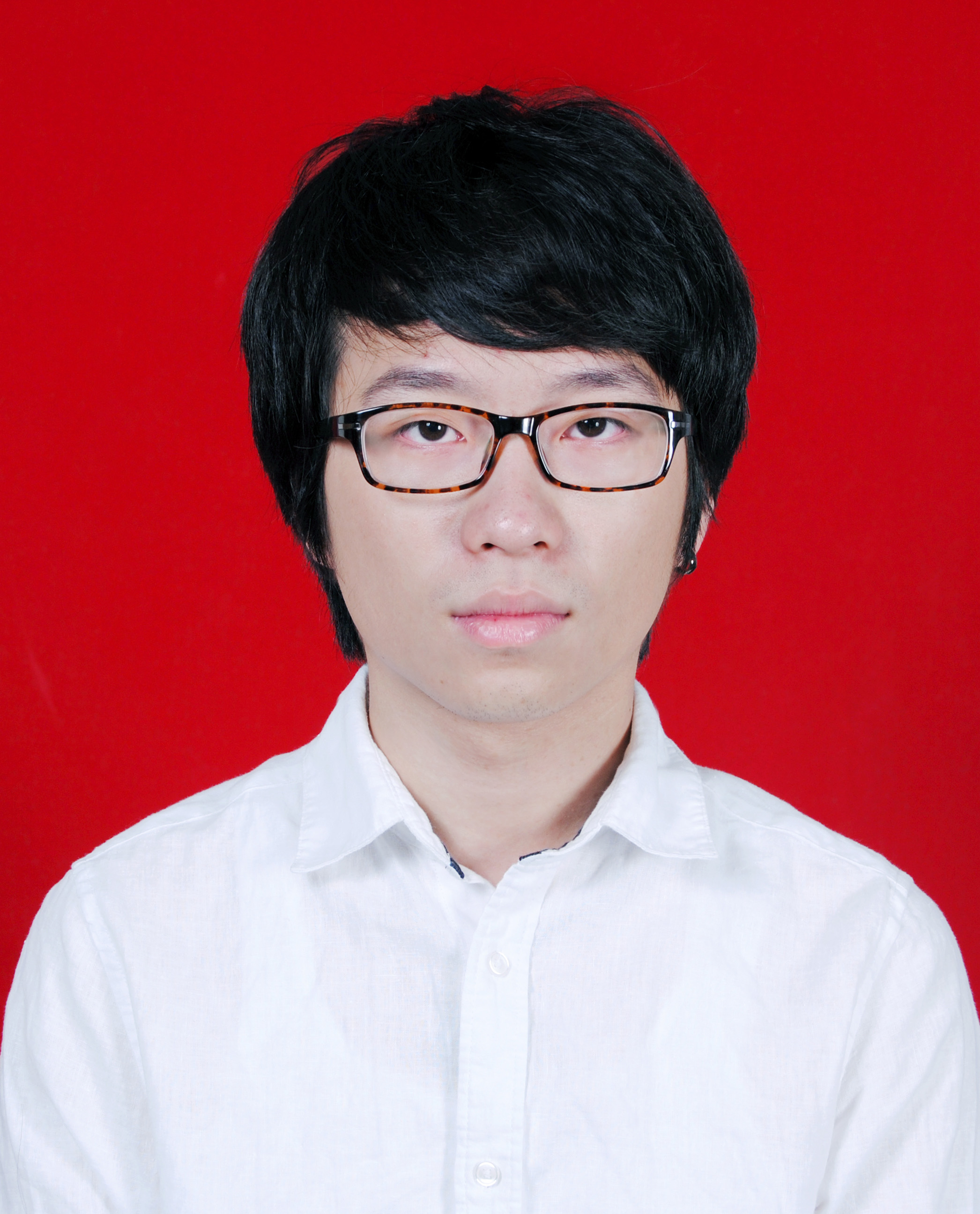}}]{Xuan Hu}
(S'16) received the B.S. degree in Electrical and Information Engineering in 2013 from Huaqiao University, Xiamen, China, and the M.S. degree in Electrical Engineering in 2015 from Arizona State University, Tempe, AZ, USA.\par
He is currently an Electrical Engineering Ph.D. student with the Erik Jonsson School of Engineering \& Computer Science at The University of Texas at Dallas, Richardson, TX, USA.\par
His current research focus is on circuit design and modeling of efficient logic and neuromorphic circuits composed of ambipolar carbon nanotubes, skyrmions, magnetic domain-wall devices, memristors, and multi-gate transistors.
\end{IEEEbiography}%

\begin{IEEEbiography}[{\includegraphics[width=1in,height=1.25in,clip,keepaspectratio]{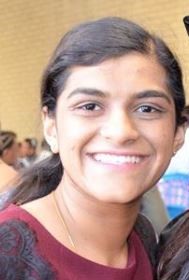}}]{Amy S. Abraham}
is a junior at the Erik Jonsson School of Engineering \& Computer Science at The University of Texas at Dallas, majoring in Electrical Engineering and minoring in Computer Science. Her current research focus is on circuit design and modeling of efficient logic circuits composed of ambipolar carbon nanotubes.
\end{IEEEbiography}	
	
\begin{IEEEbiography}[{\includegraphics[width=1in,height=1.25in,clip,keepaspectratio]{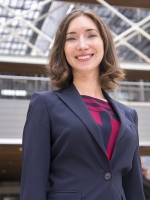}}]{Jean Anne C. Incorvia}
(M'11) received the B.A. degree in physics from the University of California, Berkeley, Berkeley, CA, USA in 2008, and the A.M. and Ph.D. degrees in physics from Harvard University, Cambridge, MA, USA in 2012 and 2015, respectively. \par
She was a postdoctoral research associate at Stanford University, Stanford, CA, USA from 2015-2017, and joined the University of Texas at Austin Department of Electrical and Computer Engineering as an assistant professor in 2017. Her technical contributions include magnetic compute-in-memory devices and circuits, and device and materials research in novel materials including low dimensional materials and emerging memories. She received a 2020 National Science Foundation CAREER award as well as the 2020 IEEE Magnetics Society Early Career Award.

\end{IEEEbiography}

\begin{IEEEbiography}[{\includegraphics[width=1in,height=1.25in,clip,keepaspectratio]{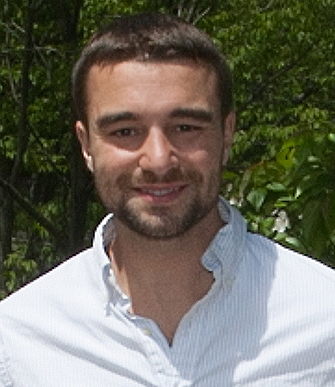}}]{Joseph S. Friedman}
(S'09-M'14-SM'19) received the A.B. and B.E. degrees from Dartmouth College, Hanover, NH, USA, in 2009, and the M.S. and Ph.D. degrees in electrical \& computer engineering from Northwestern University, Evanston, IL, USA, in 2010 and 2014, respectively.\par
He joined The University of Texas at Dallas, Richardson, TX, USA, in 2016, where he is currently an Assistant Professor of electrical \& computer engineering and the Director of the NeuroSpinCompute Laboratory. From 2014 to 2016, he was a Centre National de la Recherche Scientifique Research Associate with the Institut d'Electronique Fondamentale, Universit\'{e} Paris-Sud, Orsay, France. He has also been a Summer Faculty Fellow at the U.S. Air Force Research Laboratory, Rome, NY, USA, a Visiting Professor at Politecnico di Torino, Turin, Italy, a Guest Scientist at RWTH Aachen University, Aachen, Germany, and worked on logic design automation as an intern at Intel Corporation, Santa Clara, CA, USA.\par
Dr. Friedman is a member of the editorial board of \textit{Microelectronics Journal}, the technical program committees of DAC, DATE, SPIE Spintronics, NANOARCH, GLSVSI, ICECS, and VLSI-SoC, the review committee of ISCAS, and the IEEE Circuits \& Systems Society Nanoelectronics and Gigascale Systems Technical Committee. He has been a member of the organizing committee of VLSI-SoC 2020, NANOARCH 2019, and DCAS 2018. He has also been awarded a Fulbright Postdoctoral Fellowship. His research interests include the invention and design of novel logical and neuromorphic computing paradigms based on nanoscale and quantum mechanical phenomena, with particular emphasis on spintronics.

\end{IEEEbiography}

\end{document}